\documentclass[12pt,aps,prb]{revtex4}
\usepackage{epsfig}
\usepackage{amsmath}

\normalsize

\def\bk{{\bf k}}

\def\b0{{\bf 0}}

\def\bra{\langle}
\def\ket{\rangle}
\def\up{\uparrow}
\def\down{\downarrow}

\def\eps{\epsilon}

\def\om{\omega}
\def\Om{\Omega}
\def\sg{\sigma}

\begin{document}


\title{Competition of Fermi surface symmetry breaking \\ 
       and superconductivity}

\author{Hiroyuki Yamase and Walter Metzner}

\affiliation{Max-Planck-Institute for Solid State Research, 
 D-70569 Stuttgart, Germany}

\date{\small\today}


\begin{abstract}
We analyze a mean-field model of electrons on a square lattice with
two types of interaction: forward scattering favoring a $d$-wave
Pomeranchuk instability and a BCS pairing interaction driving 
$d$-wave superconductivity. 
Tuning the interaction parameters a rich variety of phase diagrams 
is obtained.
If the BCS interaction is not too strong, Fermi surface symmetry 
breaking is stabilized around van Hove filling, and coexists with
superconductivity at low temperatures. 
For pure forward scattering Fermi surface symmetry breaking 
occurs typically via a first order transition at low 
temperatures. The presence of superconductivity
reduces the first order character of this transition and,
if strong enough, can turn it into a continuous one.
This gives rise to a quantum critical point within the
superconducting phase.
The superconducting gap tends to suppress Fermi surface 
symmetry breaking. 
For a relatively strong BCS interaction, Fermi surface symmetry
breaking can be limited to intermediate temperatures, or can be
suppressed completely by pairing.
\\
\noindent
\mbox{PACS: 71.10.Fd, 71.10.Pm, 74.20.Mn}
\end{abstract}


\maketitle

\section{Introduction}

Usually the Fermi surface of an interacting electron systems respects
the point-group symmetry of the underlying crystal lattice. 
However, electron-electron interactions may also lead to Fermi
surface deformations which break the orientational symmetry 
spontaneously.
From a Fermi liquid viewpoint this can happen via a \lq\lq Pomeranchuk
instability'', that is, when Pomeranchuk's stability condition
\cite{Pom} for the forward scattering interactions is violated.

Interactions favoring a symmetry-breaking Fermi surface deformation
with a $d$-wave order parameter, where the surface expands along the
$k_x$-axis and shrinks along the $k_y$-axis (or vice versa), are
present in the $t$-$J$,\cite{YK1} Hubbard,\cite{HM,GKW} and extended
Hubbard\cite{VV} model on a square lattice.
These models therefore exhibit enhanced \lq\lq nematic'' correlations,
which also appear in the context of fluctuating stripe 
order.\cite{KFE}
Signatures for such correlations have been observed in various
cuprate superconductors.\cite{KBX}
In particular, they provide a natural explanation for the
relatively strong in-plane anisotropy observed in the magnetic 
excitation spectrum of $\rm YBa_2Cu_3O_y$.\cite{HPX,YM}

Fermi surface symmetry breaking competes with superconductivity.
In the $t$-$J$ model the $d$-wave Fermi surface deformation instability 
is overwhelmed by $d$-wave pairing. This is indicated by slave-boson 
mean-field theory\cite{YK1} and has been confirmed recently by a
variational Monte Carlo calculation.\cite{EMG}
However, enhanced nematic correlations remain.\cite{Yam}
The competition of superconductivity and Fermi surface symmetry
breaking is more delicate in the two-dimensional Hubbard model.
Renormalization group calculations in the symmetric phase suggest
that the superconducting instability is always stronger than the 
Pomeranchuk instability,\cite{HSR} but 
these calculations do not exclude the possibility of \emph{coexistence} 
of the two competing order parameters in the symmetry-broken phase.
Indeed, coexistence of $d$-wave superconductivity and $d$-wave 
Fermi surface symmetry breaking has been obtained near van Hove filling 
from a weak coupling perturbation expansion for the symmetry-broken
ground state of the Hubbard model.\cite{NM}

To elucidate the interplay and competition of Fermi surface symmetry
breaking and superconductivity in a more general setting, and to 
classify possible scenarios, we analyze in the present work a mean-field 
model allowing for both instabilities with a tunable strength for each.
The model describes itinerant electrons on a square lattice with 
two types of interaction: a reduced BCS interaction driving $d$-wave
superconductivity and a purely forward scattering interaction driving 
$d$-wave Fermi surface symmetry breaking.

The properties of the mean-field model without BCS interaction, where
the electrons interact only via forward scattering (\lq\lq f-model''), have 
been clarified already earlier.\cite{KKC,KCOK,YOM}
The main results can be summarized as follows.
Fermi surface symmetry-breaking occurs below a transition temperature
$T_c$ which forms a dome-shaped line as a function of the chemical
potential $\mu$, with a maximal $T_c$ near van Hove filling.
\cite{KKC,KCOK}
The phase transition is usually first order at the edges of the
transition line, and always second order around its center.
\cite{KKC,KCOK,YOM}
The $d$-wave compressibility of the Fermi surface is however strongly
enhanced even near the first order transition down to zero temperature.
\cite{YOM}
Adding a uniform repulsion to the forward scattering interaction, the
two tricritical points at the ends of the second order transition line
are shifted to lower temperatures. 
For a favorable choice of hopping and interaction parameters one of the 
first order edges can be replaced completely by a second order transition
line, leading to a quantum critical point.\cite{YOM}
Fluctuations at and near the quantum critical point destroy fermionic
quasi-particle excitations, leading to non-Fermi liquid behavior.
\cite{MRA,DM}

Adding an attractive $d$-wave BCS interaction to the f-model leads to
a variety of qualitatively distinct phase diagrams, depending on the
interaction strength.
If the BCS interaction is not too strong, Fermi surface symmetry 
breaking is stabilized around van Hove filling, and coexists with
superconductivity at low temperatures. 
In the presence of a pairing gap it is easier to realize Fermi surface 
symmetry breaking via a continuous phase transition at low temperatures 
than without.
In particular, a quantum critical point connecting superconducting 
phases with and without Fermi surface symmetry breaking at zero 
temperature is obtained for a suitable choice of interactions.
For a relatively strong BCS interaction, Fermi surface symmetry
breaking can be limited to intermediate temperatures, or can be
suppressed completely by pairing.

The article is structured as follows.
In Sec.~II we introduce the mean-field model and describe the 
self-consistency equations for the order parameters.
The phase diagrams and other results are presented in Sec.~III.
A conclusion follows in Sec.~IV.

\section{Mean-field model}

We analyze itinerant electrons on a square lattice interacting
via forward scattering and a reduced BCS interaction, described
by a Hamiltonian of the form
\begin{equation}
 H = \sum_{\bk} \eps_{\bk}^0 n_{\bk} + H_I^f + H_I^c \; ,
\end{equation}
where $n_{\bk} = \sum_{\sg} n_{\bk\sg}$ counts the spin-summed
number of electrons with momentum $\bk$. 
The kinetic energy is due to hopping between nearest and 
next-nearest neighbors on a square lattice, 
leading to the bare dispersion relation
\begin{equation}
 \eps_{\bk}^{0}= -2t (\cos k_{x}+\cos k_{y}) 
 - 4t'\cos k_{x} \cos k_{y} \; . 
\end{equation}
The forward scattering interaction reads
\begin{equation}
 H_I^f = 
 \frac{1}{2L} \sum_{\bk,\bk'} f_{\bk\bk'} \, n_{\bk} n_{\bk'} \; ,
\end{equation}
where $L$ is the number of lattice sites, and the function 
$f_{\bk\bk'}$ has the form
\begin{equation}
 f_{\bk\bk'} = u - g_f \, d_{\bk} d_{\bk'} \; ,
\end{equation}
with coupling constants $u \geq 0$ and $g_f \geq 0$, and a function 
$d_{\bk}$ with $d_{x^2-y^2}$-wave symmetry such as 
$d_{\bk} = \cos k_x - \cos k_y$.
This ansatz mimics the structure of the effective interaction in 
the forward scattering channel as obtained for the $t$-$J$\cite{YK1} 
and Hubbard\cite{HM} model.
The uniform term originates directly from the repulsion between 
electrons and suppresses the (uniform) electronic compressibility 
of the system. 
The $d$-wave term enhances the $d$-wave compressibility and drives 
spontaneous Fermi surface symmetry breaking.
In the Hubbard model it is generated by fluctuations, while in 
the $t$-$J$ model the nearest neighbor interaction contributes 
directly to a $d$-wave attraction in the forward scattering channel.

The BCS interaction has the form
\begin{equation}
 H_I^c = \frac{1}{L} \sum_{\bk,\bk'} V_{\bk\bk'} \,
 c^{\dag}_{\bk\up} c^{\dag}_{-\bk\down} 
 c_{-\bk'\down} c_{\bk'\up} \; .
\end{equation}
It is a reduced interaction in the sense that it contributes only
in the Cooper channel, that is, when the total momentum of the 
interacting particles vanishes.
For the matrix element $V_{\bk\bk'}$ we choose a separable $d$-wave 
attraction
\begin{equation}
 V_{\bk\bk'} = - g_c \, d_{\bk} d_{\bk'}
\end{equation}
with $g_c \geq 0$, which corresponds to the dominant term in the
Cooper channel for the two-dimensional Hubbard and $t$-$J$ model.

Inserting $n_{\bk} = \bra n_{\bk} \ket + \delta n_{\bk}$ into
$H_I^f$, and $c^{\dag}_{\bk\up} c^{\dag}_{-\bk\down} =
\bra c^{\dag}_{\bk\up} c^{\dag}_{-\bk\down} \ket + 
\delta (c^{\dag}_{\bk\up} c^{\dag}_{-\bk\down})$ into $H_I^c$,
and neglecting terms quadratic in the fluctuations, one obtains
the mean-field Hamiltonian
\begin{equation}
 H_{\rm MF} = \sum_{\bk} \Big[ 
 \eps_{\bk} \, n_{\bk} +
 (\Delta_{\bk} \, c^{\dag}_{\bk\up} c^{\dag}_{-\bk\down} + h.c.) 
 - \frac{\delta\eps_{\bk}}{2} \bra n_{\bk} \ket
 - \Delta_{\bk} \, 
 \bra c^{\dag}_{\bk\up} c^{\dag}_{-\bk\down} \ket \Big] \; .
\end{equation}
Here $\eps_{\bk} = \eps^0_{\bk} + \delta\eps_{\bk}$ is a 
renormalized dispersion relation, which is shifted with 
respect to the bare dispersion by
$\delta\eps_{\bk} = 
 L^{-1} \sum_{\bk'} f_{\bk\bk'} \bra n_{\bk'} \ket =
 un + \eta \, d_{\bk} \,$, 
where $n = L^{-1} \sum_{\bk} \bra n_{\bk} \ket$ is the
average particle density, and
\begin{equation}
 \eta = - \frac{g_f}{L} 
 \sum_{\bk} d_{\bk} \bra n_{\bk} \ket
\end{equation}
is our order parameter for Fermi surface symmetry breaking.
It vanishes as long as the momentum distribution function 
$\bra n_{\bk} \ket$ respects the symmetry of the square 
lattice.
The superconducting gap function is given by
$\Delta_{\bk} = \frac{1}{L} \sum_{\bk'} V_{\bk\bk'} \,
 \bra c_{-\bk'\down} c_{\bk'\up} \ket =
 \Delta \, d_{\bk} \,$, 
where
\begin{equation}
 \Delta = - \frac{g_c}{L} 
 \sum_{\bk} d_{\bk} \bra c_{-\bk\down} c_{\bk\up} \ket \; .
\end{equation}
For the reduced interactions $H_I^f$ and $H_I^c$ the 
mean-field decoupling is exact in the thermodynamic limit.
Feynman diagrams describing contributions beyond mean-field
theory have zero measure for $L \to \infty$.

The mean-field Hamiltonian is quadratic in the Fermi operators 
and can be diagonalized by a Bogoliubov transformation.
For the grand canonical potential per lattice site,
$\om = L^{-1} \Om$, we obtain
\begin{equation}
 \om(\eta,\Delta) = 
 - \frac{2}{\beta L} \sum_{\bk} \log[2\cosh(\beta E_{\bk}/2)] +
 \frac{\eta^2}{2g_f} + \frac{|\Delta|^2}{g_c} + 
 un - \frac{un^2}{2} - \mu \; ,
\end{equation}
where $\beta$ is the inverse temperature, 
$E_{\bk} = (\xi_{\bk}^2 + |\Delta_{\bk}|^2)^{1/2}$, and
$\xi_{\bk} = \eps_{\bk} - \mu$.
The stationarity conditions $\partial\om/\partial\eta = 0$
and $\partial\om/\partial\Delta = 0$ yield the 
self-consistency equations for the order parameters
\begin{equation}
 \eta = \frac{g_f}{L} \sum_{\bk} d_{\bk} \,
 \frac{\xi_{\bk}}{E_{\bk}} \, \tanh\frac{\beta E_{\bk}}{2}
\end{equation}
and
\begin{equation} 
 \Delta = \frac{g_c}{L} \sum_{\bk} d_{\bk} \,
 \frac{\Delta_{\bk}}{2E_{\bk}} \, \tanh\frac{\beta E_{\bk}}{2}
 \; ,
\end{equation}
respectively. The condition $\partial\om/\partial n = 0$ 
(at fixed $\mu$) yields the equation determining the density
\begin{equation}
 n = 1 - \frac{1}{L} \sum_{\bk}
 \frac{\xi_{\bk}}{E_{\bk}} \, \tanh\frac{\beta E_{\bk}}{2}
 \; .
\end{equation}

\section{Results}

We now show results obtained from a numerical solution of the
mean-field equations. For the ratio of hopping amlitudes we
choose $t'/t = -1/6$. The bare dispersion $\eps_{\bk}^0$ has 
saddle points at $\bk = (\pi,0)$, $(0,\pi)$, leading to a 
logarithmic van Hove singularity in the bare density of states
at $\eps = - 4t' = -2t/3$. All the results presented in the 
figures are for $u = 0$ (no uniform contribution to forward
scattering), but we will discuss the effects of a finite 
$u$ in the text.
In the following we set $t=1$, that is, all results with
dimension of energy are in units of $t$.

In Fig.~1 we show the transition temperature $T_f(\mu)$ for
Fermi surface symmetry breaking in the absence of pairing
($\Delta = 0$) for $g_f = 1$, and the critical
temperature for superconductivity $T_c(\mu)$ in the absence
of Fermi surface symmetry breaking ($\eta = 0$) for 
various choices of $g_c$. 
As discussed in detail in Refs.~\onlinecite{KCOK} and
\onlinecite{YOM}, 
a symmetry-broken Fermi surface is stabilized below a 
dome-shaped transition line, with a maximal transition
temperature near van Hove filling. The transition is first
order at the edges of the transition line and second order
around its center. 
The critical temperature for superconductivity is also
maximal near van Hove filling, but $T_c(\mu)$ remains
finite for any $\mu$ (as long as the band is partially
filled), and the transition is always of second order.
Near van Hove filling the transition temperatures $T_f$
and $T_c$ are of the same order of magnitude for $g_c$
slightly above $g_f = 1$.
Note that in the weak coupling limit one would obtain
$T_f \ll T_c$ for comparable $g_f$ and $g_c$, since
$\log(1/T_f) \propto g_f^{-1}$ for $g_f \to 0$, \cite{YOM} 
while $\log(1/T_c) \propto g_c^{-1/2}$ for $g_c \to 0$
at van Hove filling,
due to the logarithmic divergence of the density of states, 
and the additional logarithm in the Cooper channel,
as can be seen from the gap equation (12).

We now discuss results for the full mean-field model, 
allowing also for coexistence of the two order parameters.
In Fig.~2 we show the low temperature region of the phase
diagram in the $(\mu,T)$-plane for $g_f = 1$ and a 
relatively weak BCS coupling, $g_c = 0.7$.
Fermi surface symmetry breaking suppresses $T_c$ and
remains essentially unaffected by the (relativeley small)
superconducting gap. The suppression of $T_c$ occurs
since Fermi surface symmetry breaking splits the van Hove
singularity, reducing thus the density of states at the 
Fermi level. However, superconductivity cannot be
eliminated completely, since a logarithmic Cooper 
singularity survives for any reflection invariant Fermi 
surface. The phase diagram thus exhibits three types of
first order transitions between phases with a symmetric
and a symmetry-broken Fermi surface: between two normal
states, between a superconductor and a normal state, and
between two superconducting states. 
Continuing the (then metastable) phase with a symmetric 
Fermi surface beyond the first order transition line leads 
to a diverging $d$-wave compressibility at the ficticious 
second order transition line \lq\lq $T_f^{\rm 2nd}$'' also 
shown in the plot.

For larger $g_c$ the energy scale for superconductivity
(gap and $T_c$) increases, and effects of the superconducting
gap on Fermi surface symmetry breaking become more pronounced, 
see Fig.~3 (here $g_c = 0.9$). In particular the first
order transition line $T_f(\mu)$ is shifted toward the
center of the symmetry-broken region, and approaches the
ficticious second order line \lq\lq $T_f^{\rm 2nd}$''. 
In Fig.~4 we show the $\Delta$-dependence of the \lq\lq reduced'' 
Landau energy 
$\om(\Delta) = \om[\eta_{\rm min}(\Delta),\Delta] - 
\om[\eta_{\rm min}(0),0]$ for two points in the phase
diagram which are close to each other, but on opposite
sides of the first order transition between a superconducting
state with a symmetric Fermi surface and a normal state with
Fermi surface symmetry breaking; $\eta_{\min}(\Delta)$
minimizes $\om(\eta,\Delta)$ for fixed $\Delta$.
Note that $\eta_{\min}$ is zero for large $\Delta$;
the kink in $\om(\Delta)$ is due to the discontinuous onset 
of $\eta$ for small $\Delta$.
The $\mu$-dependence of the order parameters $\eta$ and 
$\Delta$ is shown for various temperatures in Fig.~5.
The jump of $\eta$ at the first order transition induces a
counter jump of $\Delta$.
For high $T$, superconductivity is suppressed completely 
by Fermi surface symmetry breaking (Fig.~5c), while for 
lower temperatures coexistence of the order parameters 
$\Delta$ and $\eta$ is realized (Figs.~5a and 5b).
The temperature dependence of the order parameters is shown
for $\mu = -0.7$ (near van Hove filling) in Fig.~6. 
The increasing superconducting gap $\Delta$ leads to a
decrease of $\eta$ upon lowering the temperature below 
$T_c$. Superconductivity smears the single
particle states over an energy range of order $\Delta$,
and thus suppresses the energy gain from a Fermi surface
deformation.
Fermi surface symmetry breaking is thus suppressed by
the superconducting gap.

Although the system is not critical at the first order 
transition from a symmetric to a symmetry-broken Fermi 
surface, it is close to criticality in the sense that the 
$d$-wave compressibility $\kappa_d$ is strongly enhanced 
by the forward scattering interaction.
For the case of pure forward scattering (f-model) this
was shown already in Ref.~\onlinecite{YOM}.
In the presence of superconductivity with a gap function
$\Delta_{\bk}$, the $d$-wave compressibility is given by
\begin{equation}
 \kappa_d = \frac{\kappa_d^0}{1 - g_f \, \kappa_d^0} \; ,
\end{equation}
where
\begin{equation}
 \kappa_d^0 = \frac{1}{L} \sum_{\bk} d_{\bk}^2 \,
 \left[
 \frac{\beta \xi_{\bk}^2}{2 E_{\bk}^2} \,
 \frac{1}{\left(\cosh\frac{\beta E_{\bk}}{2}\right)^2} +
 \frac{|\Delta_{\bk}|^2}{E_{\bk}^3} \,
 \tanh\frac{\beta E_{\bk}}{2} \right]
\end{equation}
is the $d$-wave compressibility in the superconducting
state in the absence of forward scattering ($g_f = 0$).
The enhancement of $\kappa_d$ due to $g_f$ is thus given
by the \lq\lq Stoner factor'' $S = (1 - g_f \, \kappa_d^0)^{-1}$.
In Fig.~7 we plot the inverse Stoner factor $S^{-1}$ along
the right first order transition line (approached from the
symmetric phase) up to the tricritical temperature 
$T_f^{\rm tri}$ for various choices of $g_c$.
It becomes clear that $S$ is enhanced significantly by 
superconductivity at low temperatures. 
In particular, for $g_c = 0.9$ the system is very close
to criticality.

Enhancing $g_c$ beyond $g_c = 0.9$, the first order transition 
lines between the states with symmetric and symmetry-broken 
Fermi surfaces are successively replaced by a continuous phase 
transition. 
In Fig.~8 we show the phase diagram for $g_c = 1$. Here Fermi
surface symmetry breaking occurs via a continuous transition
at the lowest temperatures, well below $T_c$.
In particular, there is a continuous quantum phase transition 
at $T = 0$.
The first order lines are connected to continuous transition
lines both at the high and low temperature ends. 
The low temperature ends are tricritical points, where the
quadratic and the quartic coefficient of the reduced Landau 
energy $\om(\eta) = \om[\eta,\Delta_{\rm min}(\eta)]$ both 
vanish.
By contrast, at the high temperature ends the quartic 
coefficient of $\om(\eta)$ jumps from a negative to a 
positive value. This discontinuity is due to the onset of 
$\Delta$ below $T_c$.
Note that the high temperature ends are close to the 
tricritical points found for smaller $g_c$, such that a
small jump of the quartic coefficient can turn its sign. 
For $g_c = 1.12$ the first order transition has disappeared
completely from the phase diagram (see Fig.~9), and the 
transition between symmetric and symmetry-broken Fermi surfaces
is always continuous.
The transition lines for Fermi surface symmetry breaking and
superconductivity intersect in tetracritical points, where
both quadratic coefficients of $\om(\eta,\Delta)$ vanish.
Enhancing $g_c$ further leads to a progressive suppression of
Fermi surface symmetry breaking, in particular at lower
temperatures, where the superconducting gap is getting large.
For $g_c = 1.2$, Fermi surface symmetry breaking is eliminated
completely by superconductivity at low $T$, while it still 
survives in a small region at intermediate temperatures, see
Fig.~10.
For even larger $g_c$ the region with a symmetry-broken Fermi
surface shrinks further until it disappears completely from
the phase diagram.

Adding a uniform contribution $u > 0$ to $f_{\bk\bk'}$, Eq.~(4),
leads to a suppression of first order transitions into a phase
with a symmetry-broken Fermi surface, making thus continuous
transitions easier. 
This trend was already observed and explained in detail for the 
case of pure forward scattering.\cite{YOM}
For small $g_c$, the tricritical points are shifted to lower
temperatures by a finite $u$, and the first order transition 
line moves closer to the ficticious second order transition.
The gradual replacement of the first order line by a second
order for increasing $g_c$ is accelerated for $u > 0$.
For example, for $g_f = 1$, $u = 10$, and $g_c = 0.9$ the 
phase diagram looks qualitatively as the one in Fig.~9, with
Fermi surface symmetry breaking always occuring via a 
continuous transition.

The (effective) interaction resulting from the Hubbard or 
$t$-$J$ model contains also an $s$-wave component in the 
Cooper channel. 
In case of coexistence of superconductivity with a $d$-wave
Fermi surface deformation, this leads to a small $s$-wave
contribution to the gap function $\Delta_{\bk}$, in addition
to the dominant $d$-wave term.\cite{YK1,NM}

\section{Conclusions}

We have solved a mean-field model for itinerant electrons
moving on a square lattice with two types of interactions:
an interaction in the forward scattering channel favoring
a $d$-wave shaped symmetry-breaking Fermi surface deformation 
and a reduced BCS interaction with $d$-wave symmetry.
Making different choices for the interaction parameters,
a rich variety of possible phase diagrams has been found.

For pure forward scattering Fermi surface symmetry breaking 
occurs typically via a first order transition at low 
temperatures.\cite{KCOK,YOM} The presence of superconductivity
reduces the first order character of this transition and,
if strong enough, can turn it into a continuous one.
This gives rise to a quantum critical point within the
superconducting phase.
The superconducting gap tends to suppress Fermi surface 
symmetry breaking. For a certain choice of parameters one
finds reentrant behavior, where Fermi surface symmetry
breaking is stabilized at intermediate temperatures, while 
it is suppressed by the pairing gap at low temperatures.
If superconductivity is too strong, Fermi surface symmetry
breaking disappears completely from the phase diagram.

In microscopic models the relative strength of forward 
scattering and pairing interactions is determined by the 
microscopic interactions. 
In the $t$-$J$ model slave-boson mean-field\cite{YK1} and 
variational Monte Carlo\cite{EMG} calculations show that 
pairing prevents Fermi surface symmetry breaking, but there 
are strongly enhanced correlations indicating that the model 
is close to a $d$-wave Pomeranchuk instability.\cite{Yam}
This corresponds to the case of a relatively large $g_c$ 
in our phenomenological mean-field model.
In the weakly interacting Hubbard model coexistence of
superconductivity and Fermi surface symmetry breaking
has been found around van Hove filling at $T=0$ within
second order perturbation theory.\cite{NM} 
The available numerical results indicate that Fermi surface 
symmetry breaking occurs via a continuous transition in
this case, as in the phase diagrams in Figs.~8 or 9.

It would clearly be interesting to analyze how order parameter 
fluctuations modify the mean-field results. 
A renormalization group calculation by Vojta {\it et al.}\cite{VZS}
suggests that a quantum critical point for orientational
symmetry breaking in a $d$-wave superconductor is 
destabilized by fluctuations, leading possibly to a first
order transition.

\acknowledgments
We thank Marijana Kir\'{c}an for a critical reading of the manuscript
and for valuable comments.




\vfill\eject


\begin{figure}[htb]
\center
\epsfig{file=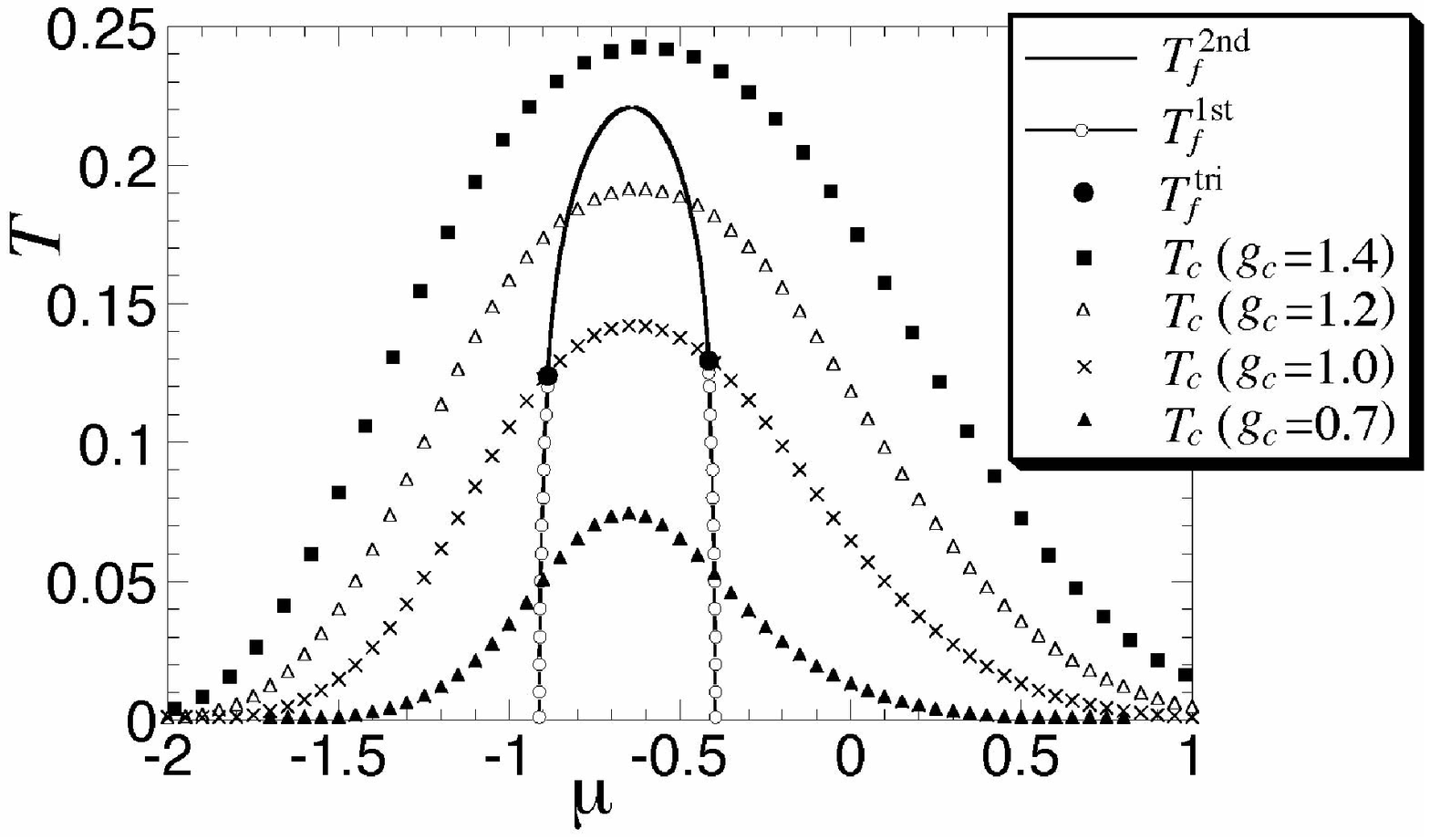,width=8cm}
\caption{Critical temperature $T_f(\mu)$ for Fermi surface
 symmetry breaking in the absence of superconductivity
 ($\Delta = 0$) for $g_f = 1$ and critical temperature
 for superconductivity $T_c(\mu)$ in the absence of Fermi
 surface symmetry breaking ($\eta = 0$) for various choices 
 of $g_c$.
 The superconducting transition is always of second order.
 Fermi surface symmetry breaking occurs via a first order
 transition at temperatures below the tricritical points 
 $T_f^{\rm tri}$, and via a second order transition above.} 
\end{figure}

\begin{figure}[htb]
\center
\epsfig{file=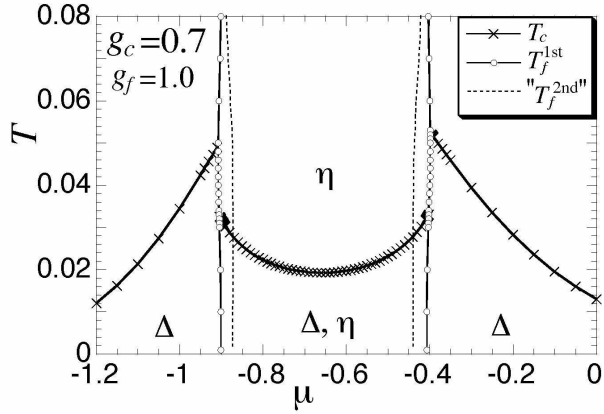,width=8cm}
\caption{Phase diagram in the $(\mu,T)$-plane for $g_f = 1$
 and $g_c = 0.7$. The symbols $\eta$ and $\Delta$ indicate
 which order parameters are finite in the various regions
 confined by the transition temperatures.
 Fermi surface symmetry breaking occurs via
 a first order transition in the temperature range shown in 
 the plot. Continuing the (then metastable) phase with a
 symmetric Fermi surface beyond the first order transition
 line leads to a diverging $d$-wave compressibility at the
 ficticious second order transition line \lq\lq $T_f^{\rm 2nd}$''
 also shown in the plot.}
\end{figure}

\begin{figure}[htb]
\center
\epsfig{file=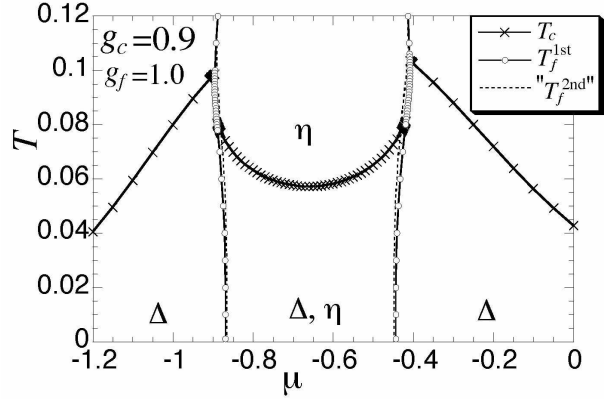,width=8cm}
\caption{Phase diagram in the $(\mu,T)$-plane for $g_f = 1$
 and $g_c = 0.9$. The first order transition line for Fermi
 surface symmetry breaking is very close to the ficticious
 second order line in this case.}
\end{figure}

\begin{figure}[htb]
\center
\epsfig{file=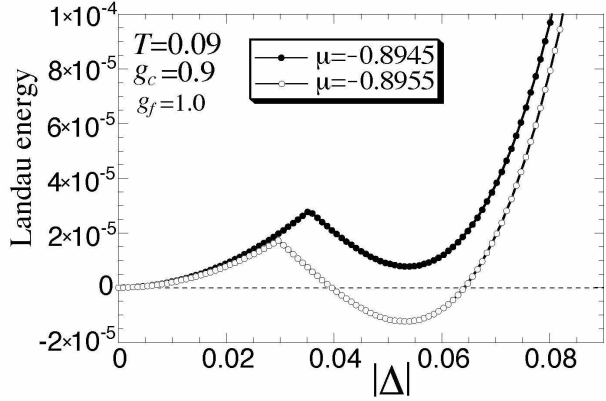,width=8cm}
\caption{Reduced Landau energy
 $\om(\Delta) = \om[\eta_{\rm min}(\Delta),\Delta] - 
 \om[\eta_{\rm min}(0),0]$, minimized with respect to $\eta$,
 as a function of $|\Delta|$.
 The interaction parameters are $g_f = 1$ and $g_c = 0.9$ as 
 in Fig.~3, the temperature is $T=0.09$. The two choices of
 $\mu$ correspond to two points close to but on opposite sides
 of the first order transition line between a superconducting
 state with a symmetric Fermi surface and a normal state with
 a $d$-wave deformed Fermi surface.}
\end{figure}

\begin{figure}[htb]
\center
\epsfig{file=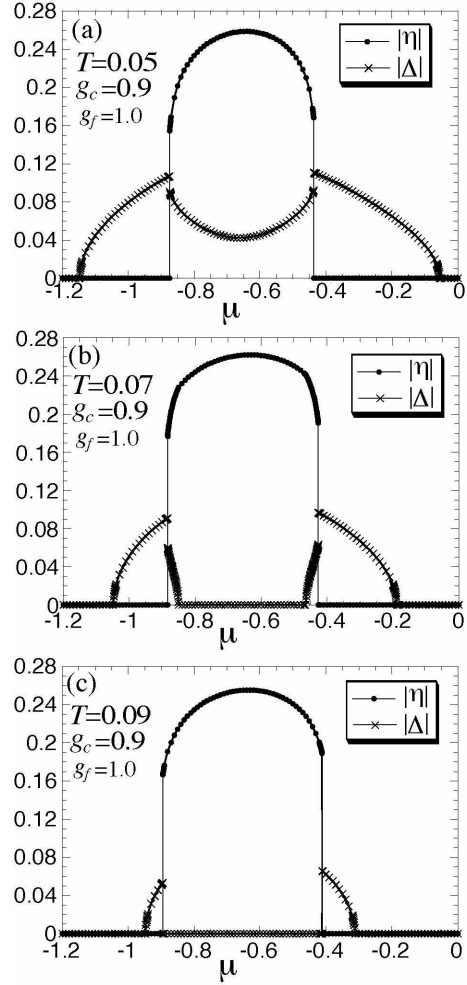,width=6cm}
\caption{Order parameters $\eta$ and $\Delta$ as a function
 of $\mu$ for various temperatures, $T = 0.05$, $0.07$, $0.09$.
 The interaction parameters are $g_f = 1$ and $g_c = 0.9$ as 
 in Fig.~3.} 
\end{figure}

\begin{figure}[ht]
\center
\epsfig{file=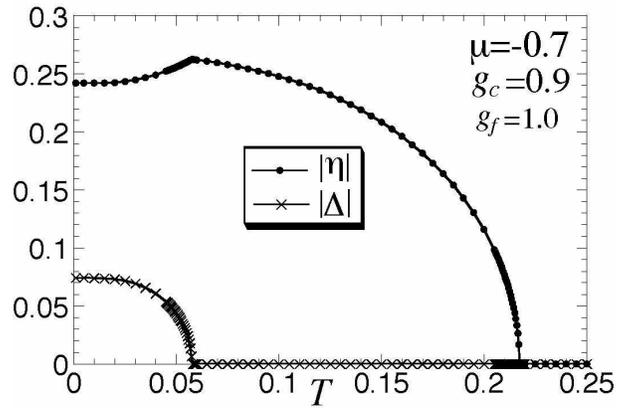,width=8cm}
\caption{Temperature dependence of the order parameters $\eta$ 
 and $\Delta$ for $\mu = -0.7$ and interaction parameters 
 as in Fig.~3.} 
\end{figure}

\begin{figure}[htb]
\center
\epsfig{file=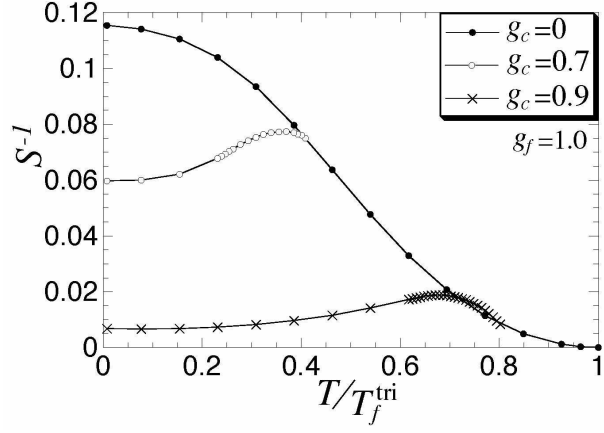,width=8cm}
\caption{Temperature dependence of the inverse Stoner 
 enhancement of the $d$-wave compressibility along the right 
 first order transition line between the phases with symmetric 
 and symmetry-broken Fermi surface. Interaction parameters are
 $g_f = 1$ and $g_c = 0$, $0.7$, $0.9$. The tricritical point
 at the high temperature end of the transition line 
 ($T_f^{\rm tri} = 0.130$) is the same in all cases.} 
\end{figure}

\begin{figure}[htb]
\center
\epsfig{file=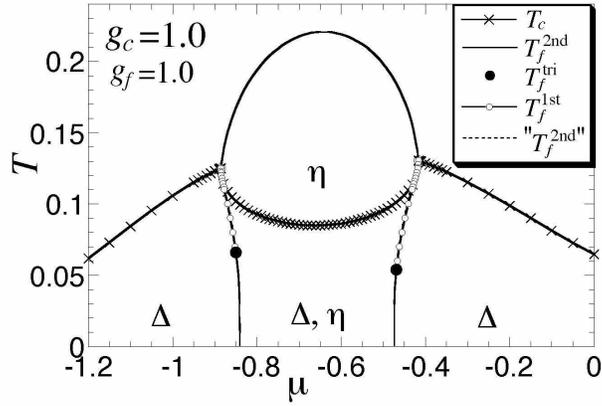,width=8cm}
\caption{Phase diagram in the $(\mu,T)$-plane for $g_f = 1$
 and $g_c = 1$. The ficticious second order transition line
 \lq\lq $T_f^{\rm 2nd}$'' is so close to the first order line
 $T_f^{\rm 1st}$ that it is hidden by the latter.}
\end{figure}

\begin{figure}[htb]
\center
\epsfig{file=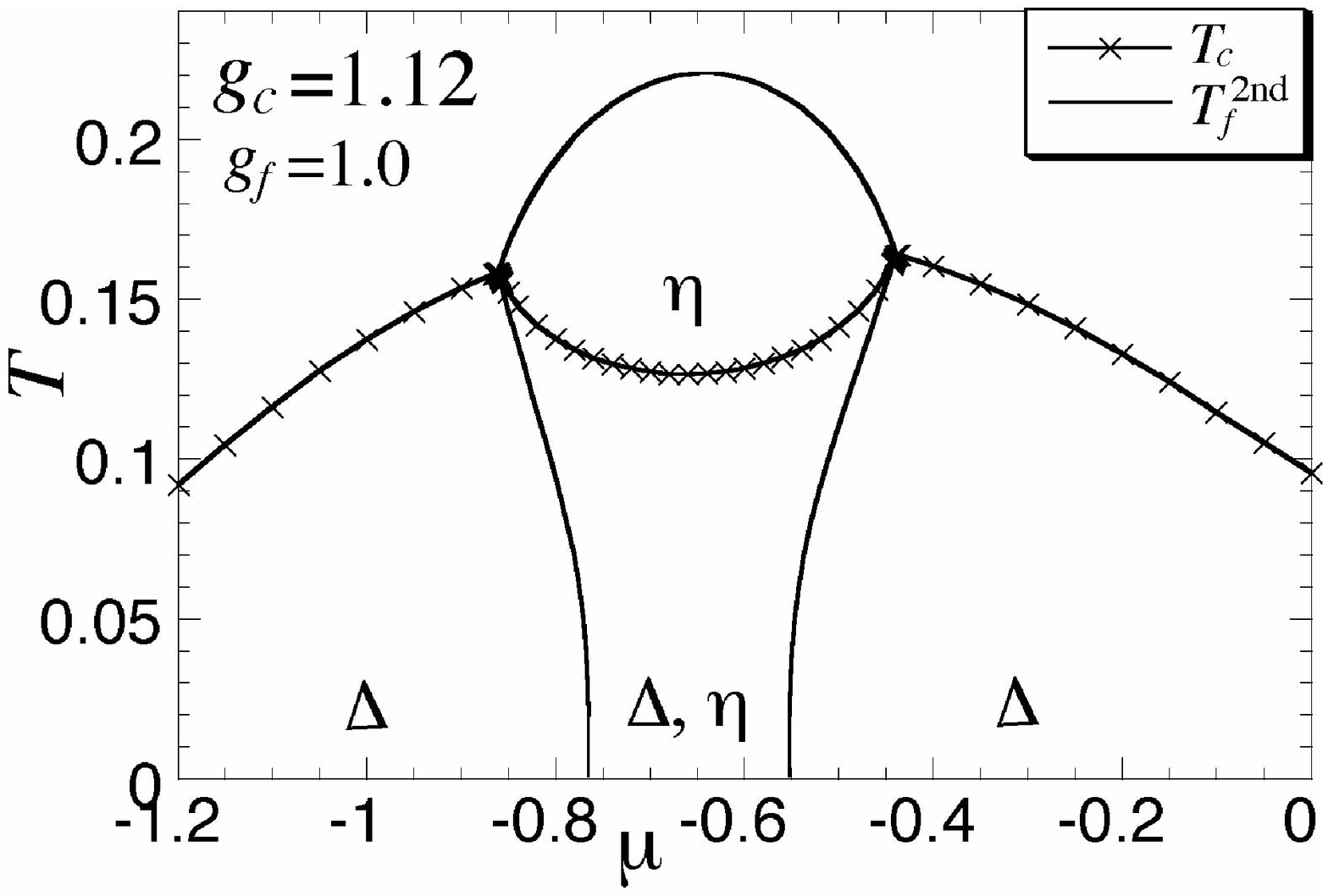,width=8cm}
\caption{Phase diagram in the $(\mu,T)$-plane for $g_f = 1$
 and $g_c = 1.12$. All transitions are of second order.}
\end{figure}

\begin{figure}[htb]
\center
\epsfig{file=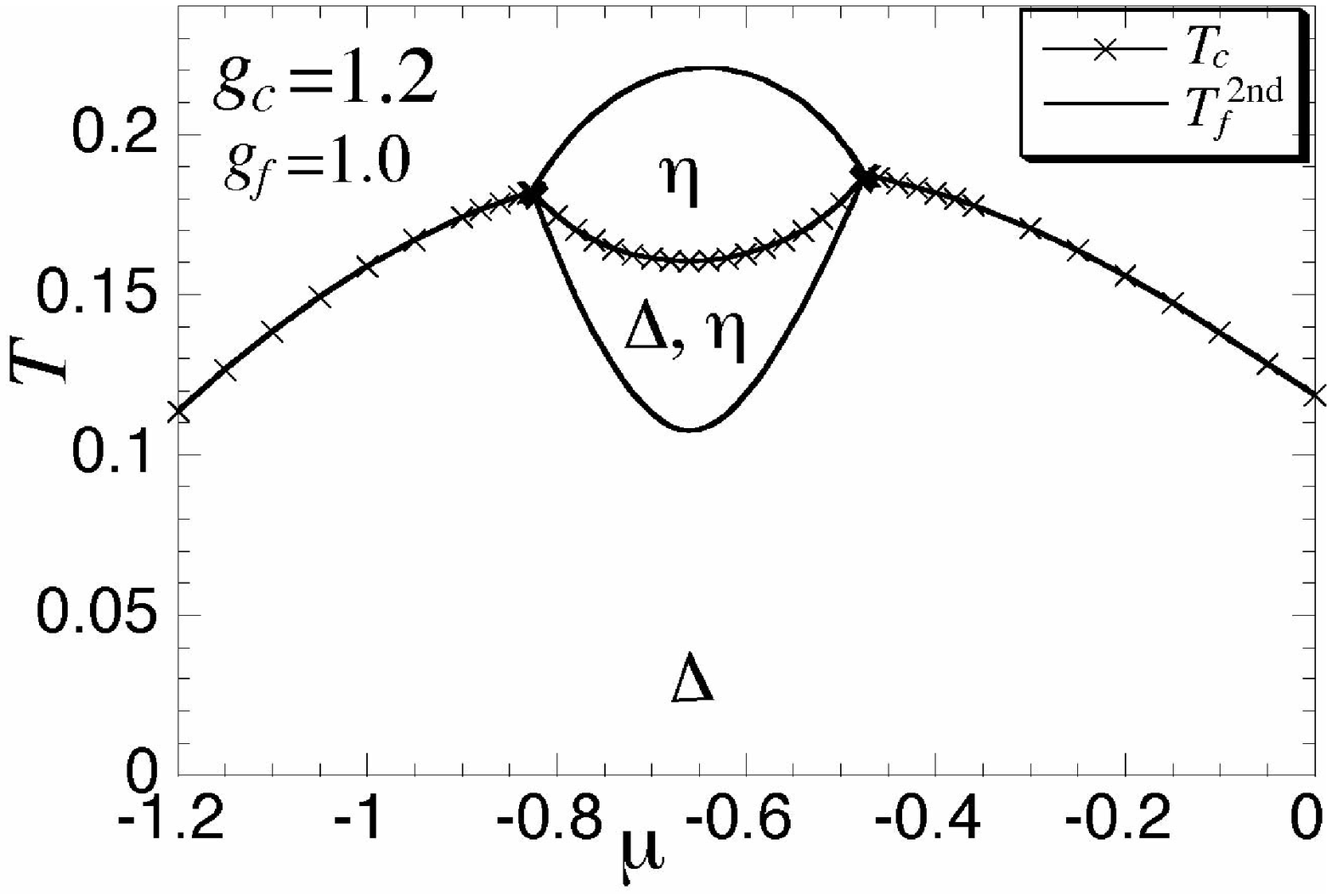,width=8cm}
\caption{Phase diagram in the $(\mu,T)$-plane for $g_f = 1$
 and $g_c = 1.2$. All transitions are of second order.}
\end{figure}

\end{document}